\documentclass[twocolumn,aps,prb,superscriptaddress,showpacs]{revtex4}
\usepackage{amsmath,amssymb}
\usepackage{graphicx} 
\usepackage{dcolumn} 
\usepackage{bm} 

\begin{document}
\draft
\title{Coherent Single Spin Source based on topological insulators}

\author{Yanxia Xing$^{1,4\dagger}$, Zhong-liu Yang$^{1}$, Qing-feng Sun$^{2,3}$, Jian Wang$^{4,\ast}$}

\affiliation{$^1$Department of Physics, Beijing Institute of
Technology, Beijing 100081, China. \\
$^2$International Center for Quantum Materials, Peking University, Beijing 100871, China\\
$^3$Collaborative Innovation Center of
Quantum Matter, Beijing, China.\\
$^4$Department of Physics and the Center of Theoretical and
Computational Physics, The University of Hong Kong, Pokfulam
Road, Hong Kong, China.
}

\begin{abstract}
We report on the injection of quantized pure spin current into quantum conductors.
In particular, we propose an on demand single spin source generated by periodically
varying the gate voltages of two quantum dots that are connected to a two dimensional
topological insulator via tunneling barriers. Due to the nature of the helical states of the topological insulator, one or several {\it spin pair}s can be pumped out per cycle giving rise to a pure quantized alternating spin current. Depending on the phase difference between two gate voltages, this device can serve as an on demand single spin emitter or single charge emitter. Again due to the helicity of the topological insulator, the single spin emitter or charge emitter is dissipationless and immune to disorders. The proposed single spin emitter can be an important building block of future spintronic devices.
\end{abstract}


\maketitle

{\sl Introduction} - Traditional electronics is based on the flow of charge where the spin of the electron is ignored. The emerging technology of spintronics\cite{spin3} will explore the spin degree of freedom such that the flow of spin, in addition to charge, will be used for electronic applications.\cite{ref1} Many applications in spintronics have been demonstrated, such as the giant magnetoresistive effect,\cite{ref2} the spin injection across a magnetic-nonmagnetic interface,\cite{ref3} and optical manipulation of spin degrees of freedom.\cite{ref4} Spin degree of freedom can also be used to process quantum information.\cite{spinbit} It is well known that quantum bit or qubit is one of the basic building blocks for quantum information science. A large variety of candidate qubit systems have been proposed such as photonic qubit\cite{photobit} and electron qubit\cite{electrbit}, and so on.

Recently, an on demand coherent single electron source has been produced experimentally\cite{charge1} and later studied theoretically.\cite{charge2} By applying ac gate voltage, periodic sequence of single electron emission and absorption on nanoseconds generate a quantized ac current. The single electron transfer between two distant quantum dots (QDs) has also been demonstrated which paves the way for single electron circuitry.\cite{exp1} This single electron source can also be used as a qubit in ballistic conductors which is an important step towards 2DEG quantum computer. The big challenge in the realization of quantum computers is to identify qubit with long coherence time. From this point, spin qubit seems to be the ideal candidate.\cite{qbit2}  This is because the spin of electron is weakly coupled to the environment compared with charge degree of freedom, the quantum coherence can be maintained at much longer time scale.\cite{spin1} It is therefore important to study the transport properties of an on demand coherent single spin source that can be used as spin qubit.

The topological insulator (TI), a new state of matter, has attracted a lot of theoretical and experimental attention.\cite{moore,kane2,zhang2} The TI has an insulating energy gap in the bulk which behaves like the general insulator, but it has exotic gapless metallic states on its edges or surfaces. The TI is first predicted in two-dimensional (2D) systems, e.g., the graphene\cite{graphene} and HgTe/CdTe quantum well \cite{HgTe} or InAs/GaSb quantum well,\cite{InAs} and has been generalized in 3D\cite{TI3D} and confirmed experimentally.\cite{hsieh} The 2D TI has the gapless helical edge states and exhibits the quantum spin Hall effect while in 3D TI the conducting state is helical surface state. This helical edge or surface states are topologically protected and are robust against all time-reversal-invariant impurities. Many interesting physical phenomena have been predicted including Majorana fermion\cite{majorana}, spin pumping or time-dependent spin injection,\cite{pump3} magneto-optical Kerr and Faraday effects,\cite{tse} and so on.

In this paper, we report on the injection of quantized pure spin current into quantum conductors. In particular, using the concept of parametric pumping,\cite{pump1,Wei,pump2} we study an on demand single spin source generated by periodically varying of gate voltage applied at two QD that are connected to a two dimensional TI via tunneling barriers. Due to the nature of helical edge state of TI, each QD will generate a fully spin polarized ac current localized near the edge of TI while the direction of the current is controlled by the phase of the gate voltage. As a result of the time reversal symmetry, either pure charge current or pure spin current can be generated depending on the phase between two gate voltages of QDs. When the phase difference is $\pi$, there is a quantized ac spin current with $n$ spin per cycle pumped out giving rise to a single spin emitter, where $n$ is an integer. When the phase different is zero, a spin unpolarized quantized ac charge current is generated with $n$ charge per cycle. We emphasize that the generated spin current has no accompanying charge current and thus is a pure ac spin source. To study the quantized spin emitter, a quantum transport theory for time-dependent pumped current using non-equilibrium Green's function method in the adiabatic regime is developed. Numerical calculations show that the quantized spin current is independent of the geometry of QD. Also, it is robust against the weak disorders.


\begin{figure}
\includegraphics[bb=4.0mm 89mm 209mm 269mm,width=8.0cm,totalheight=7.3cm, clip=]{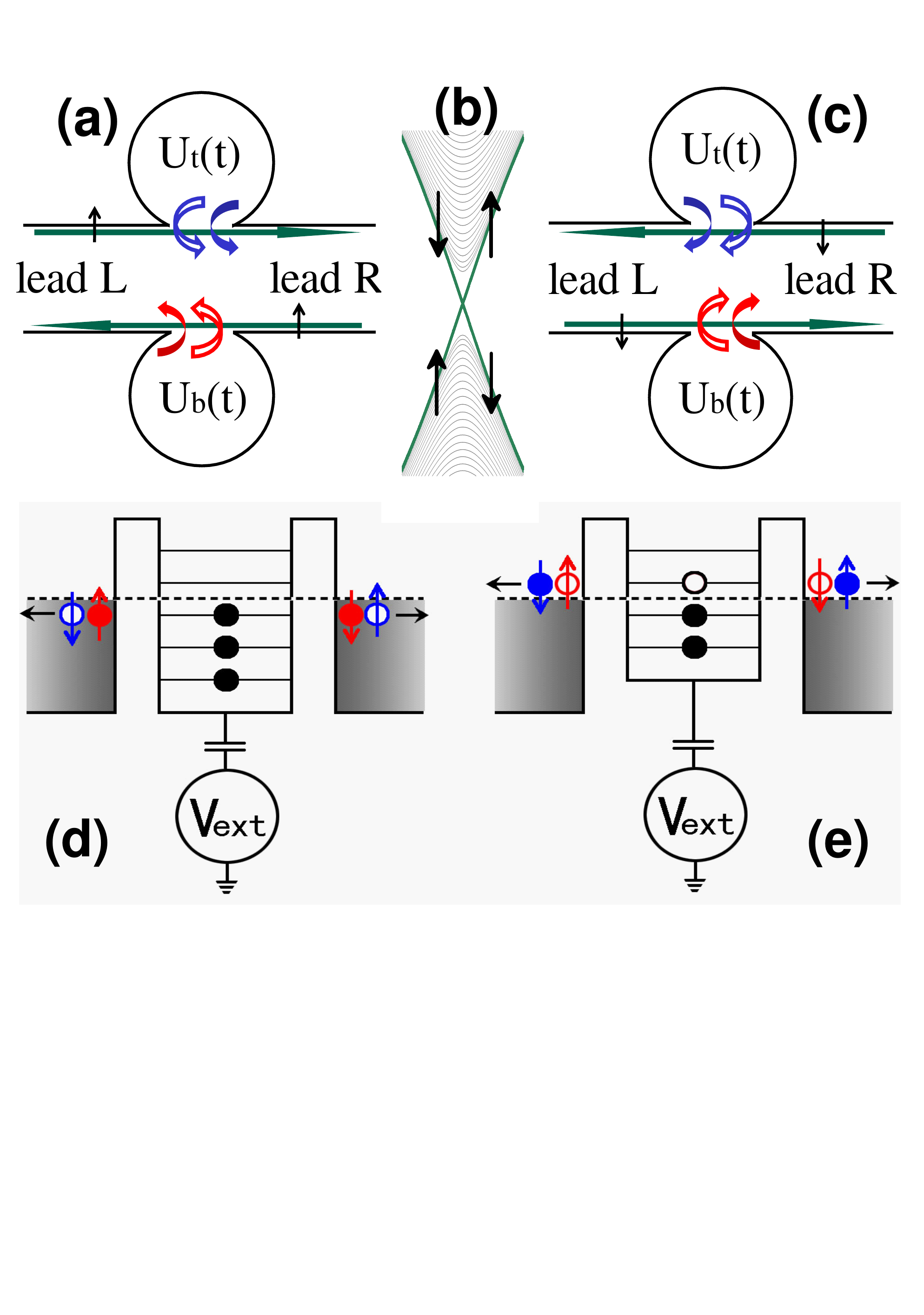}
\caption{(Color online) Panel (a) and (c): schematic plot of pumped current
in the whole period for the spin up and spin down. Panel (b): band
structure of TI. Combining panel (a) and (c), we can get Panel (d) and (e) that depict the generation of pumped pure spin
current in the first and second half period, respectively.}
\end{figure}

{\sl Model} - In a 2D TI, when the Fermi energy
is inside the energy gap, electrons can only be transported through
the unidirectional spin locked edge state. In our study, the 2D TI
ribbon is coupled with two QDs at the upper and lower edges whose
energy levels are controlled by gate voltages $V_t(t)$ and $V_b(t)$,
respectively [see Fig.1(a) or Fig.1(c)]. We assume that the amplitudes of two gate
voltages are the same but with a phase difference $\phi$. As we will
discussed below, when $\phi=0$, the system behaves like a coherent
single electron source while when $\phi=\pi$ an alternating
quantized spin current is generated, i.e., the system is an on
demand single spin emitter without accompanying electric current.

Due to the variation of the gate voltage, the current with spin up and
down [shown in Fig.1(a) and Fig.1(c), respectively] can be pumped out. In the
following discussion, we first focus on the top QD and spin up transport [see Fig.1(a)].
Note that due to the nature of the helical state there is no spin up
electron or hole from the top QD to the left lead. Assuming
that at $t=0$, $U_t=-eV_0$ is minimum, the energy level of QD is below the Fermi level and is
completely filled, i.e., filled with two electrons of opposite spin.
In the first half period, $V_{t}(t)$ decreases from $V_0$ to $-V_0$,
the energy levels of QD shifts upward from $-U_0$ to $U_0$ so that
electrons flow out of the QD and into the right lead [see the blue solid arrow in Fig.1(a)]. In the second half
period, the energy levels shift downwards from $U_0$ to $-U_0$, consequently, the holes flowing out of the QD
to the right lead [see the blue hollow arrow in Fig.1(a)].
It means under certain condition, only one electron (hole) with spin up is pumped out from the top QD
into the right lead giving rise to an alternating quantized spin
polarized charge current. In the meantime, due to the time reversal symmetry, a {\it spin
down} electron (hole) is pumped out of the QD to the left lead in the
first (second) half period [see the blue arrows in Fig.1(c)].
We emphasize that if there were no bottom QD there would be no spin
up (down) current in the left (right) lead. Now we consider the independent bottom QD in which the electrons and holes are pumped out similar to the top QD [see the red arrows in Fig.1(a) and (c)].
If the gate voltages
of the upper and lower QD are out of phase (such as $\phi_b=0,\phi_t=\pi$), the {\it spin pair} composed of electron and hole with opposite spin will be pumped into left and right leads in each half period. In this case, the net charge current is zero. However, there is a pure quantized spin current from left to right lead in the first half period and reverses its sign in the second half period as shown in Fig.1(d) and
(e). If the phase different between two gate voltages is zero, a {\it charge pair} or a {\it hole pair} with opposite spin will be pumped out leading to an alternating quantized charge current.

{\sl theory} -
The working principle of coherent single spin emitter (SSE) is based
on the 2D TI with a helical spin texture present
in momentum space.
In order to highlight the functional mechanism of
SSE and capture its salient feature, we use the modified Dirac
model with a quadratic corrections $k^2\sigma_z$, which has been shown to have
similar properties\cite{xing1} as HgTe/CdTe but with symmetric conductance and valance band.
The Hamiltonian is
given by $H(k) = [H_{\uparrow}({\bf k})+H_{\downarrow}({\bf k})]$,
where
\begin{eqnarray}
H_{\uparrow}({\bf k})&=&H^*_{\downarrow}(-{\bf k})= A({\bf
k}_x\sigma_x-{\bf k}_y\sigma_y)\nonumber \\
&+&(m+B{\bf k}\cdot{\bf k})\sigma_z+\epsilon(r)\sigma_0.\nonumber
\end{eqnarray}
Here $\sigma_{x,y,z}$ are Pauli matrices presenting the pseudospin
formed by $s,p$ orbitals and $\sigma_0$ is a unitary $2\times2$
matrix. The individual spin up Hamiltonian $H_\uparrow$ and spin
down Hamiltonian $H_\downarrow$ are time reversal symmetric to each
other, and can be calculated independently.
To carry out numerical calculation, the tight-binding Hamiltonian in
square lattice is employed, which is written as\cite{xing1,LiJianJiangHua}
\begin{eqnarray}
&&H_{\uparrow} = \sum_{\bf i} d^{\dagger}_{\bf i}
\left[\epsilon_{\bf i}\sigma_0 +(m-4t)\sigma_z\right] d_{\bf
i}+\sum_{\bf i}\nonumber
\\
&& \left[d_{\bf i}^{\dagger}
(t\sigma_z-i\frac{A}{2a}\sigma_x) d_{{\bf i}+\delta_x} +d_{\bf
i}^{\dagger} (t\sigma_z+i\frac{A}{2a}\sigma_y) d_{{\bf
i}+\delta_y}\right]+h.c. \nonumber
\end{eqnarray}
where $\epsilon_{\bf i}$ is a random on-site potential which is
uniformly distributed in the region $[-w/2,w/2]$ and ${\bf i}=({\bf
i}_x , {\bf i}_y)$ is the index of the discrete site in the unit
vectors of the square lattice with the lattice constant $a=5nm$.
$d_{\bf i}=[d_{s,{\bf i}},d_{p,{\bf i}}]^T$ with $T$ denoting
transpose, $d_{s(p),{\bf i}}$ and $d_{s(p),{\bf i}}^{\dagger}$ are
the annihilation and creation operators for s(p) orbital at site
${\bf i}$. Here $A/2a=1.35t$, $m=-0.35t$ and $t=B/a^2=27.5meV$
denote the nearest neighbor coupling strength.

To calculate the time-dependent pumped current, it is convenient to examine the
adiabatic regime. In the low frequency limit, the
system is nearly in equilibrium and the time dependent pumping
parameters are adiabatically added to the Hamiltonian.\cite{pump1} The particle
distribution in the scattering region at any instant is given by
\begin{eqnarray}
N_\sigma({\bf i},t)=-i\int \frac{dE}{2\pi}[{\bf G}^<_\sigma(E,\{\mathbf{U}(t)\})]_{\bf ii}
\end{eqnarray}
where $\sigma=\uparrow/\downarrow$ or $\pm1$ denotes the spin up and spin down,
$\mathbf{U}(t)$ is the pumping potential. Since there is no external
driving force, the left and right leads have the same Fermi energy.
From the fluctuation-dissipation theorem, ${\bf G}^<_\sigma=({\bf G}^a_\sigma-{\bf
G}^r_\sigma)f(E)$ with $f(E)$ the equilibrium distribution function for the
left and right lead, we can get total particle in the scattering
region
\begin{eqnarray}
N_\sigma(t)&=&-i\int \frac{dE}{2\pi}{\rm Tr}[({\bf G}^a_\sigma-{\bf G}^r_\sigma)f(E)]
\end{eqnarray}
where ${\bf G}^r_\sigma(E,\{\mathbf{U}(t)\})=[E{\bf I}-{\bf
H}_\sigma-{\bf\Sigma}^r_\sigma(E)-{\bf U}(t)]^{-1}$,
${\bf\Sigma}^r_\sigma={\bf\Sigma}^r_{L,\sigma}+{\bf\Sigma}^r_{R,\sigma}$ is the self energy
from the semi-infinite left and right lead, ${\bf U}$ is diagonal
matrix with ${\bf U}=\sum_{i=b/t} U_{i}{\bf\Delta}_{i}$,
where $U_{b/t}=-eV_{b/t}$ is the pumping parameter induced by gate voltage $V_{b/t}$ added in bottom or top QD, ${\bf\Delta}_{b/t}$ is the pumping potential profile which is a matrix labeling where is the pumping potential while pumping parameter $U_{b/t}$ denotes the magnitude of pumping potential.
Due to the variation of pumping parameters $U_{b/t}(t)$, the total pumped
particle current into all contacts is,\cite{Wei}
\begin{eqnarray}
dN_\sigma(t)/dt=-i\int\frac{dE}{2\pi}f(E) \sum_{i}\partial_{U_{i}}{\rm
Tr}({\bf G}^a_\sigma-{\bf G}^r_\sigma)dU_{i}/dt\label{dQdt}
\end{eqnarray}
From Dyson equation ${\bf G}^r_\sigma={\bf G}^{r,0}_\sigma+{\bf G}^{r,0}_\sigma{\bf
U}{\bf G}^{r}_\sigma$ with ${\bf G}^{r,0}_\sigma=[E{\bf I}-{\bf
H}_\sigma-{\bf\Sigma}^r_\sigma]^{-1}$, we obtain
\begin{eqnarray}
\partial_{U_{i}}{\rm Tr}[{\bf
G}^{r/a}_\sigma]={\rm Tr}[{\bf G}^{r/a}_\sigma\Delta_i {\bf
G}^{r/a}_\sigma]=-\partial_E{\rm Tr}[{\bf G}^{r/a}_\sigma{\bf \Delta}_i]
\end{eqnarray}
Note that the bold letter such as Green's functions ${\bf G}^r_\sigma$,
self energy ${\bf \Sigma^r_\sigma}$ and potential profile ${\bf \Delta}$
are all matrices which do not commute but can be rotated under the
trace operator. Using $G^a_\sigma-G^r_\sigma=iG^r_\sigma\Gamma_\sigma G^a_\sigma$ and integrating
Eq.(\ref{dQdt}) by part, we get the total instantaneous particle
current from all leads
\begin{eqnarray}
dN_\sigma/dt= \sum_i{\rm Tr}\left[\mathbf{G}^r_\sigma\left(\sum_\alpha{\bf\Gamma}_{\alpha,\sigma}\right)
\mathbf{G}^a_\sigma\mathbf{\Delta}_i\right]dU_i/dt \label{dQdt1}
\end{eqnarray}
where ${\bf \Gamma}_{\alpha,\sigma}$ is the coupling strength between
scattering region and the lead $\alpha$ for the electron with spin $\sigma$. Obviously, the above
equation gives the current partition into each lead $\alpha$.
For pumped charge and spin current, we have
\begin{eqnarray}
&&dQ_\alpha/dt=-
e\sum_\sigma dN_{\sigma,\alpha}/dt\nonumber\\
&&dS_\alpha/dt=(\hbar/2)\sum_\sigma \sigma dN_{\sigma,\alpha}/dt\label{dQdt2}
\end{eqnarray}
where $dN_{\sigma,\alpha}/dt = \sum_i{\rm Tr}\left[\mathbf{G}^r_\sigma{\bf\Gamma}_{\alpha,\sigma}
\mathbf{G}^a_\sigma\mathbf{\Delta}_i\right]dU_i/dt $.
In order to get a better understanding of the SSE, we integrate the
Eq.(\ref{dQdt1}) to get the total spin emitted in half period,
\begin{eqnarray}
S_\alpha=\frac{\hbar}{2}\int dt {\rm Tr}\sum_{\sigma}\sigma\left[\mathbf{G}^r_\sigma(E,\mathbf{U}){\bf\Gamma}_{\alpha,\sigma}
\mathbf{G}^a_\sigma(E,\mathbf{U})d\mathbf{U}\right/dt] \label{dS}
\end{eqnarray}
In the whole period, the time averaged spin current is zero, since $d\mathbf{U}$ is zero in a period.

{\sl Numerical results} -
\begin{figure}
\includegraphics[bb=0mm 0mm 195mm 145mm,width=8.5cm,totalheight=6.0cm, clip]
{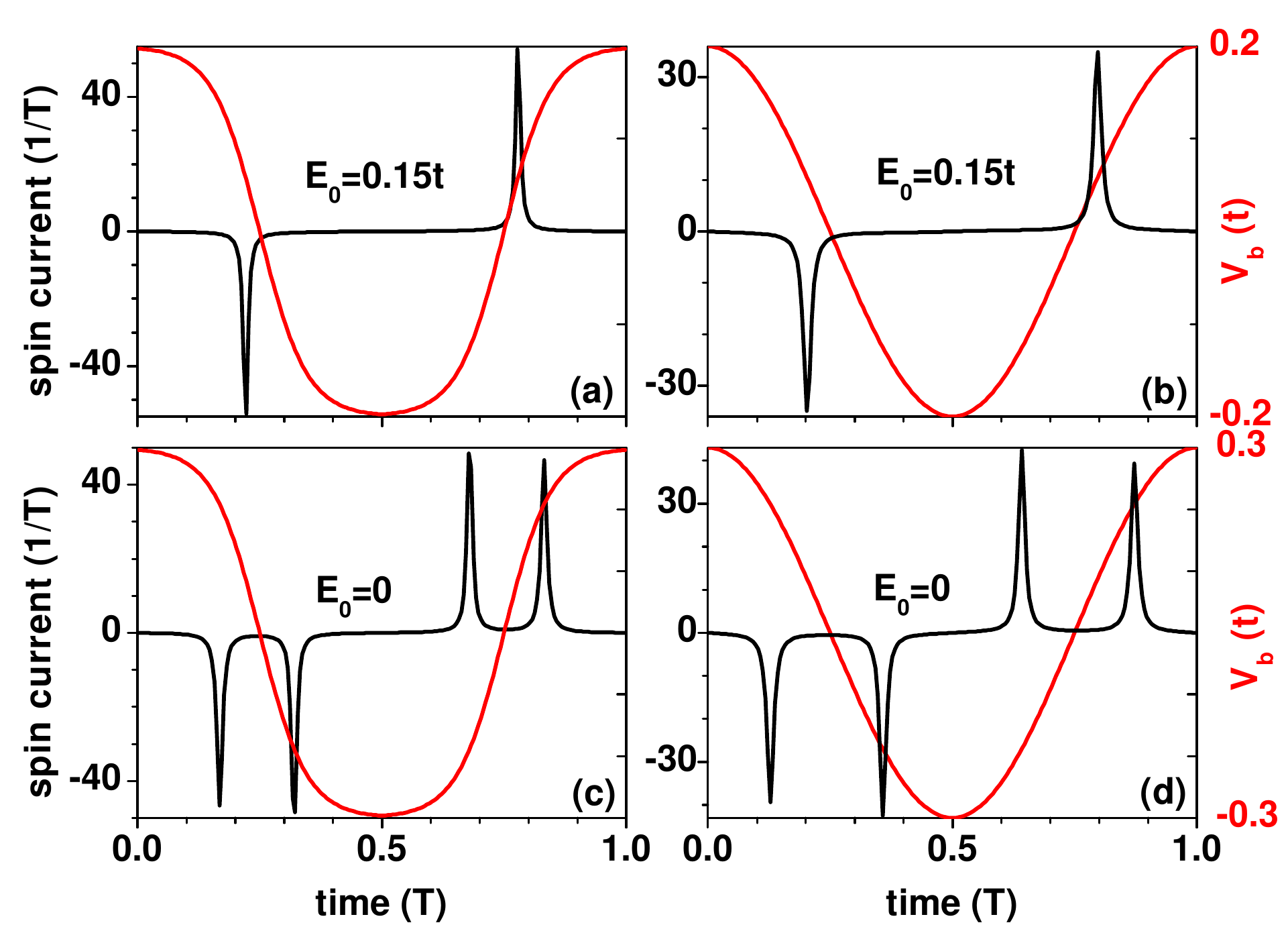} \caption{Pumped current driven by a general ac signal
[gray line in panel (a) and (c)] and harmonic signal [gray line in
panel (b) and (d)]. The red line is the profile of the gate voltage $V_b(t)$. Width of lead $W=40a$. The diameter of the circular QD is $D=21a$. The contact width of QPC $L=3a$. }
\end{figure}
In the numerical calculation, we discretize the scattering region shown in the inset of Fig.3 to obtain a tight binding Hamiltonian which can be used to construct retarded Green's function.
For convenience, we set temperature to be zero. We expect that our results remain at low temperatures, e.g., a few K. We also fix the Fermi energy of left and right leads as $E_F=0$. The side coupled QDs, modeled by modified Dirac model, have the broadening energy levels $\epsilon_{\pm n}$ that are symmetrically distributed with respect to a reference energy $E_0$. In equilibrium, i.e., $V_{b/t}=0$, $E_0=E_F=0$.
When tuning gate voltage $V_{b/t}$, $\epsilon_{\pm n}$ is shifted. For the QD, the level spacing $\Delta\epsilon$ between two nearest energy levels is around $0.2t\approx5meV$.
For this level spacing the system is in the adiabatic regime for the ac signal (a few GHz) used in most of the experiments since $\Delta\epsilon>>\hbar\omega$. In the adiabatic regime, the energy levels of QDs response to the ac gate voltage $V_{b/t}(t)$ instantaneously.
We assume that $V_{b/t}(t)=V_0\pm V(t)$ (with opposite phase), where $V_0$ is the static gate voltage that
is used to tune the reference energy $E_0$.

In Fig.2, we plot the instantaneous
pumped spin current in the whole period for the different reference energies $E_0$.
We define the positive current as the current flowing right.
It is found that the pump current peaks when the Fermi
energy sweeps through the energy levels of QD.
For $E_0=0$, when $\tau$ changes from $0$ to $T/2$, $E_F$
crosses over the two levels $\epsilon_{\pm1}$.
So, two peaks appear in Fig.2(c) and Fig.2(d). Note that due
to the coupling of the leads, the energy levels of QD $\epsilon_{\pm n}$ are slightly different
from the isolated QD $\epsilon^{iso}_{\pm n}$.
For example, $\epsilon^{iso}_{\pm 1}=\pm0.19t$ and $\epsilon_{\pm 1}=\pm0.15t$, and so on.
When $E_0=0.15t$, the $E_F$ is in line with $\epsilon_{-1}$ at $t=T/2$,
for the small amplitude $V_{b/t}$,
the Fermi energy scan over only the level of $\epsilon_{-1}$ in a half period,
then one peak appears in Fig.2(a) and Fig.2(b). Using Eq.(\ref{dS}), we can integrate the spin current and get the pumped spin pairs in a half period. For panel (a), (b), (c) and (d), the number of the {\sl spin pair} in a half
period is 1.015, 1.015, 1.957, 1.957, respectively, very close to the quantized number 1 and 2. Fig.2 also shows that various forms of ac signals such as inharmonic ac signal and harmonic signal can also generate different instantaneous currents. However, they all pump out exactly quantum {\it spin pair} in half period. So this scheme can be used as a general spin source in spintronic devices.

\begin{figure}
\includegraphics[bb=10mm 43mm 181mm 251mm,width=6.0cm,totalheight=7.3cm, clip=]
{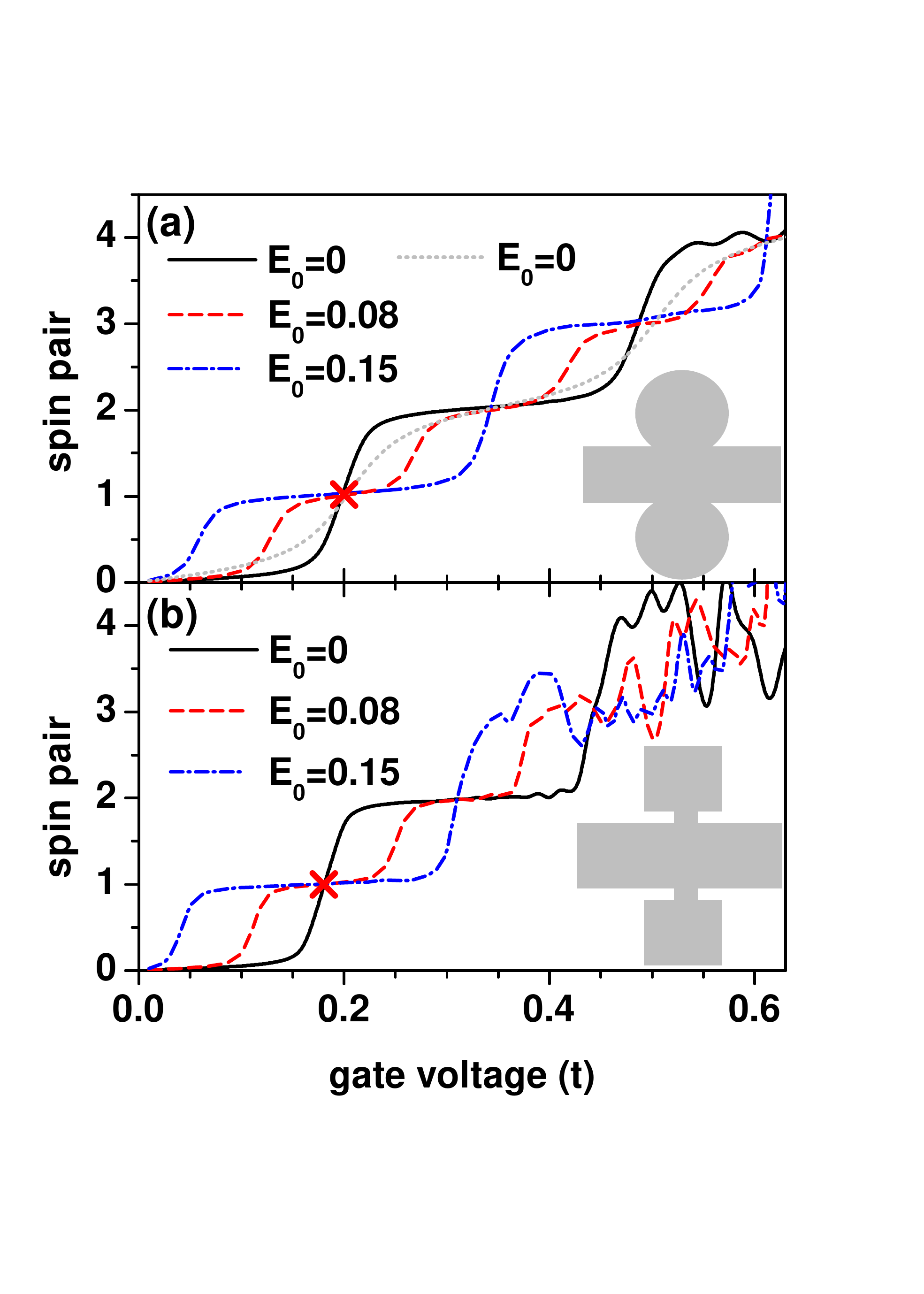} \caption{ (Color online) The number of spin pair pumped out of QD
in the first half period vs the magnitude of the ac gate voltage.
Different curves correspond to different
$E_0$. The diameter of the circular QD
is $D=21a$ and contact width $L=3a$ (thick lines) and $L=7a$ (thin
gray dotted line). The size of square QD is $20a*20a$ and contact
width $L=7a$.}
\end{figure}
In the following we will show the single spin source can be realized in
all kinds of QD with different geometric shapes. In Fig.3, we plot the
quantized {\sl spin pair}s accumulated in a half period. We consider QDs with two geometries:
circular [panel (a)] and square shaped [panel (b)]. For
different $E_0$, there are three representative configurations of
quantization: when $E_F=0$ is in line with $E_0$, $V_{b,t}$ scan through an even
number of energy levels in a half period, and number of {\it spin pair} pumped out is 0,2,4... (see black solid lines in Fig.3).
When $E_F$ is roughly in line with $\epsilon_{\pm n}$, such as $\epsilon_{-1}$ ($E_0=0.15t$),
the number of pumped {\it spin pair} is 1,3,5... (see dash dotted lines
in Fig.3), otherwise, the number of {\it spin pair} pumped out is an integer 1,2,3....
Furthermore, comparing the black solid line with the gray dotted line in Fig.3(a),
we can see that the weaker the coupling between TI and QD, the longer the quantized spin plateaus is.
This is because when the coupling is weak, the band width is small,
then it is more easier for electrons to escape completely in the voltage interval $[-V_0,V_0]$.

\begin{figure}
\includegraphics[bb=0mm 10mm 200mm 250mm,width=6.0cm,totalheight=7.3cm, clip=]{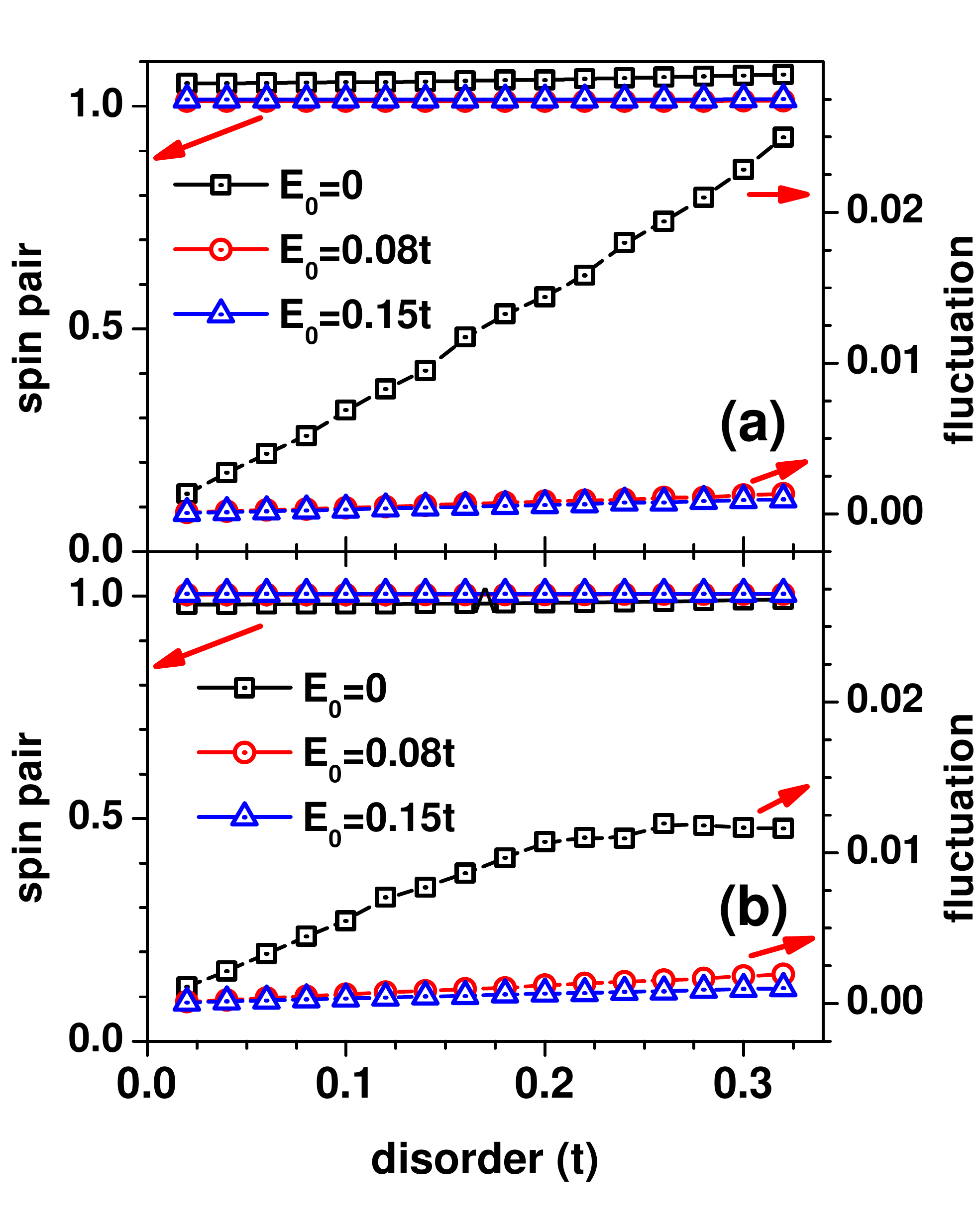}
\caption{The average number of spin pair and its fluctuation at the gate voltage marked red cross in Fig.3
vs disorder strengths for different $E_0$. The panel (a) and
(b) are corresponding to the circular and square shaped QD.
}
\end{figure}
In Fig.4, we plot the total spin on and before the first quantized plateau (the red cross in Fig.3) vs on-site disorder strengths for
different $E_0$. For $E_0=0$, the red cross (in Fig.3) is just in between two quantized plateaus. This means that the Fermi level is very close to the discrete bulk energy level and therefore a small disorder will make the helical state relax to the bulk state. For $E_0=0.08$ and 0.15, however, the Fermi level are far away from bulk energy level and the helical states are very robust. Importantly, it is shown that although fluctuation of quantized spin can be large, e.g., $E_0=0$, the
averaged values of quantized spin are hardly changed by disorder, especially for the quantized values on the quantum plateaus. This
means that this spin emitter is robust against the disorder scattering. We have not examined the effect of spin dependent disorder which are known to be present in HgTe quantum wells. This should lead to a departure of the quantized spin and charge pumping.\cite{sdis}

In conclusion, we have proposed a single spin emitter that is driven by two ac gate voltages.
Due to the helical feature of TI, an alternating pure spin current with integer number of {\it spin pair} per cycle can be generated. Importantly by tuning the phase difference between two gate voltages, either ac quantized spin current or ac quantized charge current can be pumped out. Our numerical results show that this quantized single spin emitter is robust again disorders and variation of device shapes.

{\bf Acknowledgement} We gratefully acknowledge the financial
support from from NSF-China under Grant (Nos. 11174032 and 11274364),
NBRP of China (2012CB921303), and a RGC grant
(HKU 705212P) from the Government of HKSAR.\\
$^\dagger$e-mail:xingyanxia@bit.edu.cn\\
$^\ast$e-mail: jianwang@hkusua.hku.hk.\\

\end{document}